\documentclass[a4paper, 11pt] {article}

\usepackage{latexsym}
\usepackage{graphicx}
\usepackage{epsfig}

\begin{document}
\newcommand{\eg}{{\it e.g.}}
\newcommand{\etal}{{\it et. al.}}
\newcommand{\ie}{{\it i.e.}}
\newcommand{\be}{\begin{equation}}
\newcommand{\dd}{\displaystyle}
\newcommand{\ee}{\end{equation}}
\newcommand{\bea}{\begin{eqnarray}}
\newcommand{\eea}{\end{eqnarray}}
\newcommand{\bef}{\begin{figure}}
\newcommand{\eef}{\end{figure}}
\newcommand{\bce}{\begin{center}}
\newcommand{\ece}{\end{center}}
\def\lsim{\mathrel{\rlap{\lower4pt\hbox{\hskip1pt$\sim$}}
    \raise1pt\hbox{$<$}}}         
\def\gsim{\mathrel{\rlap{\lower4pt\hbox{\hskip1pt$\sim$}}
    \raise1pt\hbox{$>$}}}         

\begin{titlepage}

\begin{center}
{\Large Effects of Quantum Vacuum Fluctuations}

\vskip .7cm

{\Large of the Electric Field on DNA Condensation}




\vskip 1cm

Alfredo Iorio$^{(a), *}$, Samik Sen$^{(b), **}$, Siddhartha Sen$^{(c,d), \dag}$ \\
$^{(a)}$ Faculty of Mathematics and Physics, Charles University of Prague \\
V Hole\v{s}ovickach 2, 180 00 Prague 8 - Czech Republic \\
$^{(b)}$  Rocwood 27, Stillorgan (Co. Dublin) - Ireland \\
$^{(c)}$ School of Mathematical Sciences, University College Dublin \\
Belfield, Dublin 4 - Ireland \\
$^{(d)}$ Indian Association for the Cultivation of Science \\
Jadavpur, Calcutta 700032 - India

\vskip 1cm

\today

\end{center}

\begin{abstract}
\noindent By assuming that not only counter-ions but DNA molecules as well are thermally distributed
according to a Boltzmann law, we propose a modified Poisson-Boltzmann equation at the classical
level as starting point to compute the effects of quantum fluctuations of the
electric field on the interaction among DNA-cation complexes. The latter are modeled here as infinite
one-dimensional wires ($\delta$-functions). Our goal is to single out such quantum-vacuum-driven interaction from
the counterion-induced and water-related interactions. We obtain a universal, frustration-free Casimir-like (codimension 2)
interaction that extensive numerical analysis show to be a good candidate to explain the formation and
stability of DNA aggregates. Such Casimir energy is computed for a variety of configurations of up to
19 DNA strands in a hexagonal array. It is found to be strongly many-body.
\end{abstract}

\vfill

\noindent $^*$ E-mail: alfredo.iorio@mff.cuni.cz

\noindent $^{**}$ E-mail: samiksen@gmail.com

\noindent $^\dag$ E-mail: sen@maths.ucd.ie

\vskip 0.5cm

\noindent PACS: 87.15.ag; 87.15.-v; 87.15.bk

\noindent Keywords: Quantum calculations; Biomolecules: structure and physical properties; Structure of aggregates

\end{titlepage}

\section{Introduction}

Dilute solutions of DNA upon the addition of specific multivalent cations have remarkable properties \cite{gelbart2000}. Initially
the DNA molecules in aqueous solution ionize with two negatively charged phosphates strands remaining, giving rise to a highly
charged anion with lineal charge density of one negative charge per $1.7$ \AA. The system, in the presence of cations,
binds them by a process known as the Oosawa-Manning (OM) condensation \cite{oosawa, manning} (for a review see, e.g., \cite{gelbart2000, Levin:2002gj, nguyen1}). When about 90 per cent of the DNA negative charge is screened and when the cations have a specific valency $+ k$,
usually\footnote{\label{specific} Counter-ions such as Spermidine${}^{4+}$, are surely seen to drive the attraction
for DNA. Less clear is the matter for $k=2$, where, e.g., ions Mg${}^{2+}$ although they OM bind to the DNA strands, are
in theory not supposed to trigger attraction \cite{kornyshev1999, KLreview}, while in experiments they appear to do that \cite{qiu2007}. We shall not address this specificity here.} $k = 3$ or $k = 4$, the DNA strands collapse to form rod-like, spheroidal and toroidal aggregates, whose size can be
experimentally controlled within limits \cite{Hud}.

There have been different approaches to study these features of DNA collapse and structure formation and much progress has been
made over the last decade. Nonetheless, many aspects remain unclear. Broadly speaking, we can identify two avenues that have been explored:

In one the details of the helicoidal charge distribution of the DNA are used to compute the electrostatic interaction between two DNA strands using a
{\it linearized} Poisson-Boltzmann (PB) (or Debye-H\"{u}ckel (DH)) approach \cite{kornyshev1999} (see also the review article \cite{KLreview}).
Within this approach the attraction between two like-sign charged but suitably oriented DNA molecules and the specificity of cations driving the attraction can be predicted. However, it was also realized that when the molecules are three or more frustration comes about (for an assembly of several DNA molecules in an hexagonal array on a triangular lattice treated within this model see, e.g., \cite{harreis2003} and also \cite{cherstvy2002-2005}). Thus, clearly, this force alone cannot account for the formation and stability of DNA aggregates.

In alternative approaches the surface of a single DNA molecule is treated as a two-dimensional complex system and statistical
mechanical arguments lead to the counterion-mediated attraction between two DNA molecules \cite{oosawabook, groenbech, Ha1997, parsegian1998}. The
key idea there is that the attraction is triggered by local correlations and thermal fluctuations not accounted for in the mean-field
PB approach. These fluctuations being those of the number or charge density of the ions, they are constrained to be classical in nature.

In a recent paper \cite{IorioSenSen} we explored the possibility for {\it quantum} vacuum fluctuations to be a viable candidate to drive the formation of DNA aggregates and to hold (``glue'') them together. In this paper we want to make explicit that analysis and shall present more details and numerical results to strengthen that hypothesis.

Our starting point is the modification of the PB classical equation to include the DNA macromolecules in the Boltzmann thermal distribution.
This way delta-function potentials appear naturally in the equation to take into account the charge distribution of the DNA. Since classical
electrostatic calculations are available and detailed \cite{kornyshev1999} and since we do not expect our one-dimensional (delta-function)
model to improve those calculations, we focus on the quantum corrections considered here via an effective action approach. When the electrostatic potential is small, as the OM condensation has taken place, we approximate our {\it modified} PB equation to obtain a {\it modified} DH equation where, as said, the DNA molecule-counterions complexes (which, for brevity, we shall often simply call ``DNA strands'') are modeled as delta function potentials (carrying information relative to the charge distribution of the strand) as to be expected by general features of
self-adjoint extension of differential (Hamiltonian) operators \cite{jackiw}. The resulting classical action is then taken as the start of a Renormalization Group (RG) type of analysis in which time-dependent \textit{quantum} fluctuations of the electric field, propagating with the speed of light in the medium (and representing short distance effects) are averaged to generate an effective potential.

The nice outcome of this procedure is that it produces precisely the Casimir energy (for a review see, e.g., \cite{parsegian2006}, \cite{Nesterenko:2005xv} and \cite{parsegian1970}) in codimension 2 (lines in three dimensions) that was obtained in \cite{Scardicchio:2005hh}, as we demonstrate. There are two scales in the model: the original DH mass (inverse length) scale $\mu$, fixed by the parameters of the system, and a new mass (inverse length) scale $M$, introduced through a process of coupling constant renormalization, which is indeed a free parameter. These scales control the range of the interaction.

This energy is calculated for an assembly of $N$ strands ($N=2,4,7,19$) in arrays in a triangular lattice and we found that:
i) it is attractive, irrespective of the charge of the DNA strands; ii) it has a range of ${\cal O} (10)$ {\AA}
in the simplest model considered; iii) the energy scale in the range is greater than $k_B T$;
iv) there are important many-body effects. We then conclude that it is plausible that this quantum relativistic force is the ``glue'' holding together the aggregate of DNA strands.

The model presented here has limits. For instance, it suggests that the DNA strands collapse to a configuration of zero separation, due to an infinitely strong (singular) attractive energy at very short distances, an instance not occurring in real aggregates. This is an indication that this force is only one part of the puzzle and the full picture needs to include more\footnote{One thing to consider {\it within} the model is that at very short distances nonlinear modifications to the Casimir energy have to play a role.}. It seems to us that the key ingredients to have the full picture
are \textit{three} mechanisms: (i) the zero-point quantum interaction, that gives the universal attraction (``glue''); (ii) the ``frustrated force'', that takes into account the detailed structure (finite size and helical architecture) and the active role of counterions (these alone seem to reproduce the helical architecture \cite{bruinsmapre}); (iii) water-related forces, that need to be included.

In Section \ref{sec2} we recall the basics of the derivation of the PB equation and then propose our modification of it to move on in Section 3 to the
quantum calculation based on the averaging-over-fluctuations method. The latter is done in some details as this should make manifest what sort of interaction we are considering here and how we derive it. In Section \ref{sec4} we present our numerical analysis of the interaction between several strands, starting with two, and discuss the outcomes and limitations of the model. The last Section is dedicated to our conclusions.

\section{The modified Poisson-Boltzmann equation}\label{sec2}

Consider the electrostatic potential $\Phi(\vec{x})$ due to a charge density at finite temperature $\rho(\vec{x}, T)$ placed in a medium with dielectric constant $\epsilon$. This obeys the Poisson equation
\begin{equation}\label{Poisson}
    \nabla^2 \Phi (\vec{x}) = - \frac{4 \pi}{\epsilon} \rho(\vec{x}, T) \;.
\end{equation}
Consider now the DNA molecule as a negatively charged rod immersed
in water at room temperature ($T \simeq 300$K) with dissolved salt
whose ions have valency $z = \pm k$, with $k = 1, 2, ...$ (for
instance, when the dissolved salt is NaCl, $k=1$, corresponding to
Na$^+$ and Cl$^-$). The charge distribution of the composite
system DNA-salt is
\begin{equation}\label{rhoMS}
\rho(\vec{x}, T) = \rho_{\rm DNA} (\vec{x}, T) + k e (n_+
(\vec{x}, T) - n_- (\vec{x}, T)) \;,
\end{equation}
where $\rho_{\rm DNA} (\vec{x})$ is the charge density of the macroion, and we take
\begin{equation}\label{nBoltz}
n_\pm (\vec{x}, T) = n_0 \exp \left( \mp \frac{k e \Phi (\vec{x})}{k_B T} \right) \;,
\end{equation}
i.e. the concentration (density) of ions follows a Boltzmann distribution.
Inserting (\ref{rhoMS}) and (\ref{nBoltz}) into (\ref{Poisson}) gives the
PB equation
\begin{equation}\label{PoissonBoltzmann}
    \nabla^2 \Phi (\vec{x}) = \frac{8 \pi}{\epsilon} k e n_0 \sinh \left( \frac{k e \Phi(\vec{x})}{k_B T} \right) \;,
\end{equation}
where only the region outside the surface of the macroion is
considered, hence $\rho_{\rm DNA} = 0$ (soon we shall re-introduce
$\rho_{\rm DNA}$). Equation (\ref{PoissonBoltzmann}) is the widely
used outcome of the mean-field theory that has been extensively
applied to the study DNA molecules immersed into aqueous media with
different salts and salt concentrations. Note that both coions
(negative) and counterions (positive) are present due to the
dissolved salt and that in experiments the counterions usually do
not come only from the dissolved salt but are added
separately\footnote{That is why, sometimes, the PB equation above
does not have the symmetric expression on the right side, $\sinh
(k)$, but rather an unbalanced form $\exp (k_1) - \exp (- k_2)$.
We shall work with the symmetric form (\ref{PoissonBoltzmann}).}.

Among the successes of the PB equation is the prediction of a
phase transition -- governed by the Manning parameter $\xi =
l_B/b$, with  $l_B = q^2 / \epsilon k_B T$ the Bjerrum length (at
room temperature $l_B \simeq 7.1${\AA}) and $1/b$ the lineal charge
density that for DNA is $b = 1.7${\AA} -- where the counterions
condense onto the DNA strands screening its negative charge to a
large extent: when $\xi > 1$ counterions stick to the charged rod to
form a DNA-counterions complex. This is the OM condensation \cite{oosawa, manning}.

Our concern is to study the interaction among DNA strands {\it
after} at least 90 per cent of the negative charge has been
screened via the OM condensation, as this is the reported critical
value for collapse. It is then reasonable to consider $\Phi$
small. Furthermore, we explicitly consider the charge distribution
$\rho_{\rm DNA}$ and demand that it obeys a Boltzmann distribution
law as for the ions (see Eq.(\ref{nBoltz}))
\begin{equation}
\rho_{\rm DNA} (\vec{x}, T) = - n^0_{\rm DNA} (\vec{x}) |q| \exp
\left(  \frac{|q| \Phi (\vec{x})}{k_B T} \right) \;,
\end{equation}
where $q < 0$ is the charge of the DNA strand with
\begin{equation}
n^0_{\rm DNA} (\vec{x}) = \sum_{i=1}^N \nu_i (z_i) \delta^2 (\vec{x}_\bot - \vec{l}_i) \;.
\end{equation}
This charge density function defines our approximations: we model the DNA strands
as infinite lines all parallel to the $z$-axis and located at
$\vec{l}_i$ in the $x-y$ plane with the coefficients $\nu_i (z_i)$
carrying information on the charge structure of the DNA strand. We
further simplify our model by taking $\nu_i (z_i) = \nu =$~constant,
$\forall i = 1, ..., N$.

We now put everything together, expand
the exponential and stop at first order to obtain
\begin{equation}\label{DHmod}
\left[ - \partial^2_z - \nabla_\bot^2 + \mu^2 + \lambda \sum_{i=1}^N \delta^{(2)} (\vec{x}_\bot - \vec{l}_i) \right] \Phi(\vec{x}) = J \;,
\end{equation}
which is a modified DH equation. Here $\mu^2 = k^2 \kappa^2$, with
\begin{equation}
\kappa^{-1} = \sqrt{\frac{\epsilon k_B T}{8 \pi e^2 n_0}}
\end{equation}
the Debye screening length,
\begin{equation}
\lambda = \frac{4 \pi \nu |q|^2}{\epsilon k_B T}
\end{equation}
and
\begin{equation}
J =  - \frac{4 \pi \nu |q|}{\epsilon}  \sum_{i=1}^N \delta^{(2)} (\vec{x}_\bot - \vec{l}_i) \;.
\end{equation}
In the following the operator in square brackets will often be indicated for brevity
as $[- \nabla^2 + V(\vec{x}_\bot)]$, with obvious notations.

Delta function potentials are not new in physics (see, e.g., \cite{albeverio}).
Their appearance in the present case can also be seen as the effect of removing points
(the locations of the DNA strands) from the domain of the
differential operator $ - \nabla^2$. This naturally leads to delta functions as a
compensation for the self-adjoint extension of such operator \cite{jackiw}.

Eq.(\ref{DHmod}) can be used to compute classical electrostatic interactions. Nonetheless, we shall not do so as such calculations are available and detailed \cite{kornyshev1999} and we do not expect our one-dimensional model to improve them. Thus, in the following Section, we shall focus on the quantum corrections.

\section{Quantum fluctuations of the electric field}\label{sec3}

Let us consider small time-dependent fluctuations: $\Phi (\vec{x}) \to \Phi (\vec{x}) + \phi (\vec{x}, t)$. $\Phi$
satisfies the modified DH equation (\ref{DHmod}) that descends from the action
\begin{equation}\label{actionPhi}
{\cal A} (\Phi) = \int d^4 x  \left( \frac{1}{2} \Phi [- \nabla^2 + V(\vec{x}_\bot)] \Phi + J \Phi \right) \;,
\end{equation}
where we use units $\hbar = c = 1$, with $c$ the velocity of light in the medium and, for the sake of clarity,
we included an integration over time $\int_0^\tau dt$ even though the functions are
time-independent. We then demand that to the fluctuation field $\phi$ as well is associated an action that is a suitable
modification of (\ref{actionPhi}), namely
\begin{equation}
\bar{{\cal A}} (\phi) = \int d^4 x \frac{1}{2} \phi (- \partial_t^2 - \nabla^2 + V(\vec{x}_\bot)) \phi \;.
\end{equation}
Note that in $\bar{{\cal A}} (\phi)$ the term with the coupling to the ``external current'' $J$ is zero because
\begin{equation}
\int d^4 x J \phi = \int d^3 x J \int_0^\tau dt \phi = 0 \;,
\end{equation}
as we impose $\int_0^\tau dt \phi = 0$ as required for fluctuating
fields. The field $\phi$, though, is a quantum field, hence the
field configurations that satisfy the classical equations (the
ones descending from $\delta \bar{{\cal A}} (\phi) = 0$) are just
on the same footing as all other field configurations\cite{feynman}. The way
to consider the effects of $\phi$ is to average these fluctuations
out to obtain an effective action ${\cal A}_{\rm eff} (\Phi)$. This is done by considering the
generating functional
\begin{eqnarray}
Z[\Phi, \phi]  & = & \int [D\Phi] e^{i {\cal A} (\Phi)} \int [D\phi] e^{i \bar{{\cal A}} (\phi)} \\
& = & \int [D\Phi] e^{- ( {\cal A} (\Phi) + {\rm corrections})}\;,
\end{eqnarray}
where we Wick rotate on the time direction $t \to i t$, and
identify ${\cal A}_{\rm eff} (\Phi) = {\cal A} (\Phi) +$
corrections.

Thus we need to compute $I = \int [D\phi] e^{- \bar{{\cal A}} (\phi)}$
that, using standard Gaussian functional integrals methods (see, e.g., \cite{ramond}),
gives $I = [\det(-\partial_t^2 - \nabla^2 + V(\vec{x}_\bot))]^{-1/2}$ or
\begin{equation}
I =  \exp \left( - \frac{1}{2} {\rm Tr} \ln (- \partial_t^2 -
\partial^2_z - \nabla_\bot^2 + V(\vec{x}_\bot)) \right) \;,
\end{equation}
where the identity $\exp((-1/2)\ln \det A) = \exp ((-1/2) {\rm Tr}
\ln A)$ was used and, as customary, the determinant and trace have
to be computed in terms of the eigenvalues of the operator\footnote{Here a squared length
$L^2$, which we set to 1, is understood to make the argument of the ``$\ln$'' and of ``$\det$'' dimensionless. Only
at the end of the computation (see Eq.(\ref{Elndet})) we shall take that into account and shall introduce the
proper scale.}. To find such eigenvalues we suppose that
\begin{equation}
\phi(\vec{x}_\bot, z, t) = \frac{1}{2 \pi}
\int_{-\infty}^{+\infty} dp e^{i p z} \phi_p (\vec{x}_\bot) e^{i
\omega t}\;,
\end{equation}
and that $(- \nabla_\bot^2 + V(\vec{x}_\bot)) \phi_p(\vec{x}_\bot) = E \phi_p(\vec{x}_\bot)$.  With this we obtain
\begin{equation}
I =  \exp \left( - \frac{1}{2} \int_{-\infty}^{+\infty} \frac{d
\omega}{2 \pi} \int_{-\infty}^{+\infty} \frac{d p}{2 \pi}
\int_0^{+\infty} dE \rho (E) \ln (\omega^2 + p^2 + E) \right) \;.
\end{equation}
Let us focus on $I_p (E) = (1/ 2\pi) \int_{-\infty}^{+\infty} d
\omega \ln (\omega^2 + E_p^2)$, where $E_p^2 = p^2 + E$. One has
that $\partial I_p(E) / \partial E = 1/ (2 E_p)$ which gives
\begin{equation}
I_p(E) = \int^E dE' \frac{\partial I_p(E')}{\partial E'} =
\frac{1}{2} \int^E dE' \frac{1}{\sqrt{E' + p^2}} = \sqrt{E + p^2}
\;,
\end{equation}
hence
\begin{equation}
I =  \exp \left( - \frac{1}{2} \int_{-\infty}^{+\infty} \frac{d
p}{2 \pi} \int_0^{+\infty} dE \rho (E) \sqrt{E + p^2} \right) \;.
\end{equation}
Thus the corrections to the classical action ${\cal A} (\Phi)$ are
${\cal E} \tau$ with the energy $\cal E$ given by
\begin{equation}\label{Ecasimir}
{\cal E} = \frac{1}{2} \int_{-\infty}^{+\infty} \frac{d p}{2 \pi} \int_0^{+\infty} dE \rho (E) \sqrt{E + p^2} \;.
\end{equation}
This energy is of the form ${\cal E} = (1/2) \sum
\omega$, i.e. the zero point Casimir energy of the system with
$\sum$ replaced by $ (1/2\pi) \int d p \int dE \rho (E)$ and
$\omega$ by $\sqrt{E + p^2}$. Noticing that the density of states
$\rho(E)$ contains information on the location $\vec{l}_i$ of the $N$ strands
present in the system (see later discussion), clearly $\partial {\cal E} / \partial \vec{l} $ is a (Casimir) force.

The mathematical problem of determining the $\cal E$ of Eq.(\ref{Ecasimir}) has been solved in the context of scalar quantum fluctuations for interacting strings \cite{Scardicchio:2005hh}. Let us give here a brief account of that derivation (see also \cite{Jaffe:2005wg}).

We first need to find the density of states $\rho(E)$. Define\footnote{Not to clutter the formulae, for the time being we shall not use any special symbol for vectors as done earlier. } $H_0 \equiv - \nabla^2 + \mu^2$ and $H = H_0 + \lambda \sum \delta^{(2)}(x - l_i) \equiv H_0 + W$. The Green's functions are such that
\begin{equation}
    (H - E) G(x,y) = \delta^{(2)} (x - y) \;,
\end{equation}
where $<x|G|y> = G(x,y)$ and similarly for $G_0(x,y)$ with $H_0$. Hence, formally we have $G = 1/(H - E)$ and $G_0 = 1/(H_0 - E)$ which
allows to set-up the Lipmann-Schwinger equation
\begin{eqnarray}
G & = & \frac{1}{H - E} = \frac{1}{H_0 - E} - \frac{1}{H_0 - E} W \frac{1}{H - E} \\
& = & G_0 - G_0 W G = G_0 - G_0 W G_0 + G_0 W G_0 W G = \cdots \\
& = & G_0 - G_0 W \frac{1}{1 + G_0 W} G_0 \;,
\end{eqnarray}
or
\begin{equation}\label{G}
    G(x,y) = G_0(x,y) + \sum_{i,j=1}^N \frac{G_0(x,l_i)G_0(l_j,y)}{\sigma \delta_{i j} - G_0(l_i,l_j)} \;,
\end{equation}
where $\sigma = - \lambda^{-1}$. Now define the energy eigenfunctions as $\psi_n(x) = <x|E_n>$ that give
\begin{equation}\label{rho}
    \rho(E) = \sum_{n =0}^\infty \int d^2 x |\psi_n(x)|^2 \delta^{(2)}(E - E_n) \;,
\end{equation}
and use
\begin{eqnarray}
    G(x,y) & = & <x| \frac{1}{H - E + i\epsilon}|y> = \sum_{n =0}^\infty \psi_n(x)\frac{1}{E_n - E + i \epsilon} \psi^*_n (y) \nonumber \\
& = & \sum_{n =0}^\infty \psi_n(x)\left( {\cal P} \frac{1}{E_n - E} + i \pi \delta (E - E_n) \right)\psi^*_n (y) \;.
\end{eqnarray}
Clearly
\begin{equation}\label{rho2}
    \rho (E) = \frac{1}{\pi} {\rm Im} \int d^2 x G(x,x) \;.
\end{equation}
We need now to prove an identity
\begin{eqnarray}
\int d^2 x G_0(x, l_i)G_0(l_j, x) & = & \int d^2 x <x| \frac{1}{H_0 - E}|l_i><l_j | \frac{1}{H_0 - E} |x>   \nonumber \\
= <l_j | \frac{1}{(H_0 - E)^2} |l_i> & = & \frac{\partial}{\partial E} <l_j | \frac{1}{H_0 - E} |l_i>  = \frac{\partial}{\partial E} G_0(l_i,l_j) \;.
\end{eqnarray}
With this and by using the expression (\ref{G}) for the propagator in the expression (\ref{rho2}) for the density of states we obtain
\begin{eqnarray}
\rho (E) & = & \rho_0 (E) + \frac{1}{\pi} \sum_{i,j=1}^N \frac{\partial G_0(l_i,l_j) / \partial E}{ \sigma \delta_{i j} - G_0(l_i,l_j)} \nonumber \\
 & = & \rho_0(E) - \frac{1}{\pi} \frac{\partial}{\partial E} \sum_{i,j=1}^N  \ln {\bar \Gamma}_{i j} \;.
\end{eqnarray}
where $\rho_0 (E) = (1/\pi) {\rm Im} \int d^2 x G_0(x,x)$ and $\bar{\Gamma}_{i j} = \sigma \delta_{i j} - G_0 (l_i,l_j)$. By dropping the $l$-independent $\rho_0 (E)$ and by using the fact that ${\bar \Gamma}_{i j}$ is diagonalizable we obtain
\begin{equation}\label{rho3}
\rho (E) = -\frac{1}{\pi} \frac{\partial}{\partial E} {\rm Tr} \ln {\bar \Gamma}_{i j} = - \frac{1}{\pi} \frac{\partial}{\partial E} \ln \det {\bar \Gamma}_{i j} \;.
\end{equation}

\begin{figure}
 \centering
  \includegraphics[height=.2\textheight]{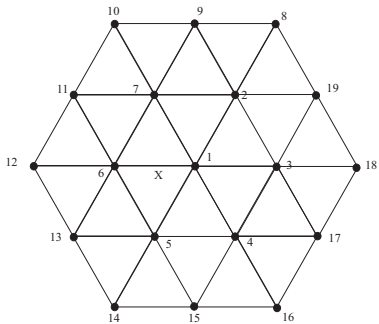}
  \caption{The configuration used for 19 DNA strands where $x$ is the distance between
  nearest neighbors. }
\label{steps}
\end{figure}

The solution for $G_0 (l_i, l_j)$, when the energy is Wick rotated to the real negative values is \cite{Scardicchio:2005hh}
\begin{equation}\label{Ko}
    G_0 (l_i, l_j) = \frac{1}{2 \pi} K_0 (\sqrt{E + \mu^2} l_{i j}) \;,
\end{equation}
where $K_0 (x)$ is the modified Bessel function of the second kind
of order zero and $l_{i j} = |l_i - l_j|$. Hence $G_0 (L) \to \infty$ when $L \to 0$. One cures this divergence by splitting $G_0(L)$ into a finite part and a divergent part and renormalizing the coupling constant $\lambda$. This process introduces, for dimensional reasons, the scale $M$ that, for stability needs to be constrained: $M < \mu$ \cite{Scardicchio:2005hh}. The result of this procedure is
\begin{eqnarray}
\sigma - G_0(L) & = & - \frac{1}{\lambda} + \frac{1}{2 \pi} \ln (M L) + \frac{1}{2 \pi} \ln \left( \frac{\sqrt{E + \mu^2}}{M} \right) \\
& \equiv &  - \frac{1}{\lambda (L)} + \frac{1}{2 \pi} \ln \left( \frac{\sqrt{E + \mu^2}}{M} \right) \\
& = & \frac{1}{2 \pi} \ln \left( \frac{\sqrt{E + \mu^2}}{M e^{2 \pi / \lambda (L)}} \right) \;,
\end{eqnarray}
where we used the fact that $K_0 (x) \sim - \ln x$ for small $x$.
We now require $\lambda \to - \infty$ when $L \to 0$ so to have a finite $\lambda_R = \lambda(L)|_{L \to 0}$ and redefine the scale
$M \to M e^{2 \pi / \lambda_R} \equiv M$ which we still require to satisfy $M < \mu$.

\begin{figure}
 \centering
  \includegraphics[height=.3\textheight]{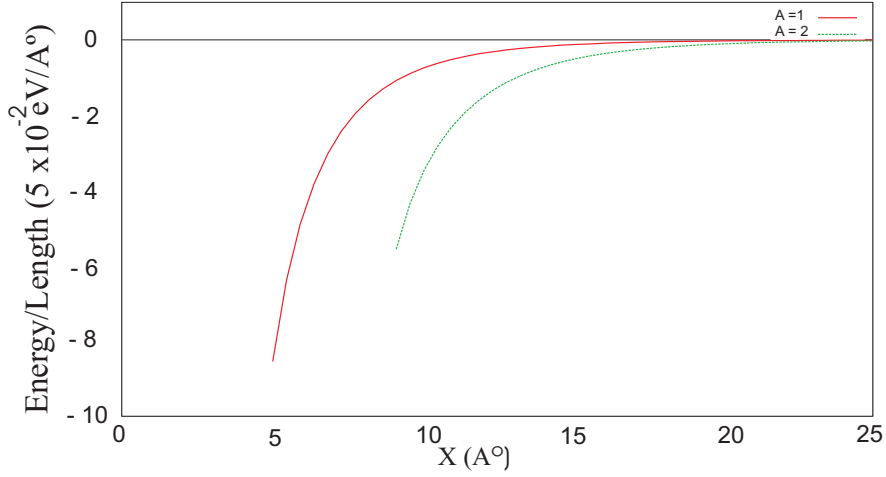}
  \caption{Energy of interaction of two DNA strands. The lower (upper) curve corresponds to $a =2$ ($a=1$). Distances $x$
  are measured in units of $1/ \mu \sim$ ${\cal O} (10)$ \AA, while lineal energy density units are estimated to be
  $5 \times 10^{-2}$~eV/\AA.}
\label{2_string_energy}
\end{figure}

Thus the diagonal terms of ${\bar \Gamma}_{i j}$ are
\begin{equation}
{\bar \Gamma}_{i i} = \frac{1}{2 \pi} \ln \left( \frac{\sqrt{E + \mu^2}}{M} \right) \equiv {\bar \Gamma} \quad \forall i = 1, ..., N \;,
\end{equation}
and represent the self-interaction. They can be obtained also by taking the asymptotic form of ${\bar \Gamma}_{i j}$
\begin{equation}\label{gammainfty}
{\bar \Gamma}_{i j} (l_{i j} \to \infty) = \bar{\Gamma} \delta_{i j} \;.
\end{equation}
These terms do not contribute to the Casimir force as they do not depend on $l_{i j}$, hence we drop them by considering
$\Gamma_{i j} \equiv {\bar \Gamma}_{i j} / {\bar \Gamma}$. Collecting all this we can write
\begin{equation}
{\cal E} = - \frac{1}{(2 \pi)^2} \int_{-\infty}^{+\infty} d p \int_0^{+\infty} dE \sqrt{E + p^2}
\frac{\partial}{\partial E} \ln \left( \det \Gamma_{i j} \right) \;,
\end{equation}
that when we integrate out the $p$ (using dimensional regularization) and partial integrate over $E$ eventually gives
\begin{equation}\label{Elndet}
{\cal E} = \frac{1}{8 \pi} \int_0^\infty dE \ln \left( \det
\Gamma_{i j} \right) \;,
\end{equation}
with
\begin{equation}
\Gamma_{i j} = \delta_{i j} -  \frac{K_0 (\sqrt{E
+ \mu^2} \; l_{i j})}{\ln(\sqrt{E + \mu^2} / M)} (1 - \delta_{i j}) \;.
\end{equation}

\begin{figure}
 \centering
  \includegraphics[height=.3\textheight]{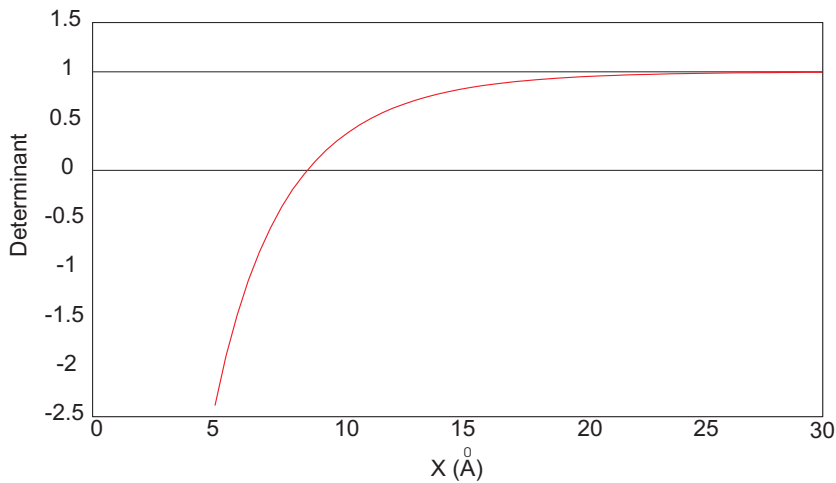}
  \caption{Determinant for the two-strand case with $a=2$.}
\label{2_K0_String_Determinants}
\end{figure}

\section{Numerical Study of the Interaction Energy}\label{sec4}

Our strategy is to study the energy of configurations of DNA strands that capture as much as possible
the symmetry of arrangements encountered in real aggregates \cite{Hud}. The first case considered is of course that
of the two-strand interaction. When then proceed to the many-body interactions with the strands sitting at the sites
of hexagonal lattices like that of Fig.~\ref{steps} which is for the maximum number of strands we were able to consider, i.e.
19 strands. We also present here results for four DNA strands sitting at the vertices of a rhombus. Thus the $x$-dependance of
${\cal E} (x)$ is that of the energy on the lattice spacing and making $x$ bigger or smaller means to expand or shrink the
aggregate size, respectively. Of course, for big enough $x$ the situation one is describing is that of a dilute solution of
DNA molecules, which is our starting point. The idealization here consist in the demanding a symmetry of arrangement even in the
dilute phase.

To render the expression (\ref{Elndet}) suitable for such a study we first numerically perform
the integral over $E$ and then plot the resulting expression ${\cal E} (x)$ where
\begin{equation}\label{ElndetNum}
{\cal E} (x) = \frac{1}{8 \pi} \ln \left( \det \gamma_{i j} (x) \right) \;,
\end{equation}
and
\begin{eqnarray}
\gamma_{i j} (x) & = & \delta_{i j} -  \frac{K_0 (\mu \; c_{i j} x)}{\ln(\mu / M)} (1 - \delta_{i j}) \\
& = & \delta_{i j} -  a K_0 (c_{i j} x) (1 - \delta_{i j}) \label{muscale}\;.
\end{eqnarray}
The relative distances $l_{i j} = c_{i j } x$ are expressed in terms of the basic lattice distance
$x$ and the numerical coefficients $c_{i j}$ take the
symmetry of the given arrangement into account. In (\ref{muscale}) we make explicit the choice $\mu = 1$ (i.e. the distances are measured in units of $\mu^{-1}$). The other scale $M$ is constrained to be positive and less than $\mu (= 1)$ and we write it as $0 < M = e^{-1/a} < 1$. In this fashion the range of $\cal E$ scales with $a$ and we present here results for $a = 1$ and $a = 2$.

The two-strand interaction energy is shown in Fig.~\ref{2_string_energy}. It is clearly attractive and finite-range. Similar attractive behaviors for the two-body interaction have been found in various models \cite{kornyshev1999, groenbech}. What we observe here is that in those models it is not clear why the interaction still needs to be attractive for more than two strands. For the Casimir energy we shall soon see that this is indeed the case, since this attraction mechanism does not suffer of any frustration.

To establish whether the magnitude and range of this attractive energy is indeed relevant for the case of DNA aggregates we need to move, as said, to the many-body case. Before doing so we need to first consider that the distances are measured in units of $1 / \mu \sim {\cal O} (10)$ \AA. This gives a range of attraction that can be adjusted by fixing the free parameter $a$ to fit the typical distances reached within the aggregates. For the hexagonally packed toroidal condensates such distances range between~\cite{Hud} 18~\AA~and 28~\AA, values clearly compatible with the range we obtain for the many-body interaction (see later).

\begin{figure}
 \centering
  \includegraphics[height=.3\textheight]{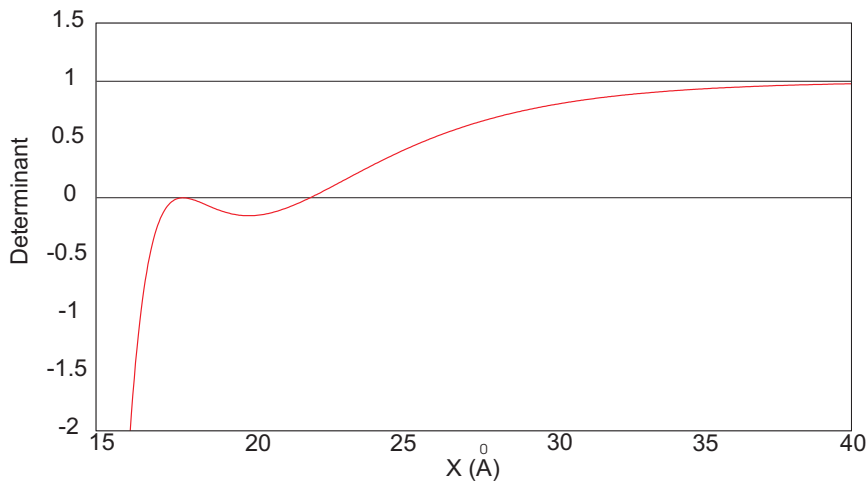}
  \caption{Determinant for the 19-strand case with $a=2$. Note that only the value $\bar{x} \sim 22$\AA~is the one to consider as the singular
  point.}
\label{2K0_19_String_Determinants}
\end{figure}

To compare this quantum energy with thermal energy we write the integral in Eq.(\ref{Elndet}) in a dimensionless fashion so that the factor in front is $\hbar c \mu^2 \sim 5 \times 10$~eV/\AA, where $c \sim 10^8 m s^{-1}$. Hence, as the values of the integrals we obtain are ${\cal O}(10^{-3})$, the unit for the energy per length (in \AA) is estimated to be ${\cal E} \sim 5 \times 10^{-2}$~eV/\AA. Thermal fluctuations, as computed for instance in \cite{groenbech}, give for the two-body interaction a maximum value of ${\cal E} \sim 5 \times 10^{-3}$~eV/\AA~at a separation distance of 10 \AA. At this distance our interaction for the two-body case gives $5 \times 10^{-2}$~eV/\AA~(for $a=1$) and $2 \times 10^{-1}$~eV/\AA~(for $a=2$), i.e. a result that is between one and two orders of magnitude stronger. For the many-body case, the case of importance for the aggregates, this factor grows enormously but we cannot trust our approximations for $x$ too close to the singular value of the logarithm, say $\bar{x}$. It is clear, though, that at the distances of relevance this quantum energy is stronger (or much stronger) than thermal energy.

Let us look more closely to the singularity of the energy. We are able to numerically evaluate this $\bar{x}$ for the various cases by plotting the determinant that of course shows no such singularity as can be seen in Fig.~\ref{2_K0_String_Determinants} ($\bar{x} \sim 8$\AA) and Fig.~\ref{2K0_19_String_Determinants} ($\bar{x} \sim 22$\AA) for the two-strand case and for the 19-strand case of Fig.~\ref{steps}, respectively. That singularity means that if only the Casimir force were present the strands would collapse to zero separation, an instance that does not occur in the real case because of other forces (not considered here) such as the electrostatic force that for more than two strands will give a net repulsive effect. Another important factor at such short distances is of course the finite size of DNA strands that have a transverse length (radius of the cylinder) of $10$\AA. We take the distance $\bar x$ as the limit of validity of our approximations.

\begin{figure}
 \centering
  \includegraphics[height=.3\textheight]{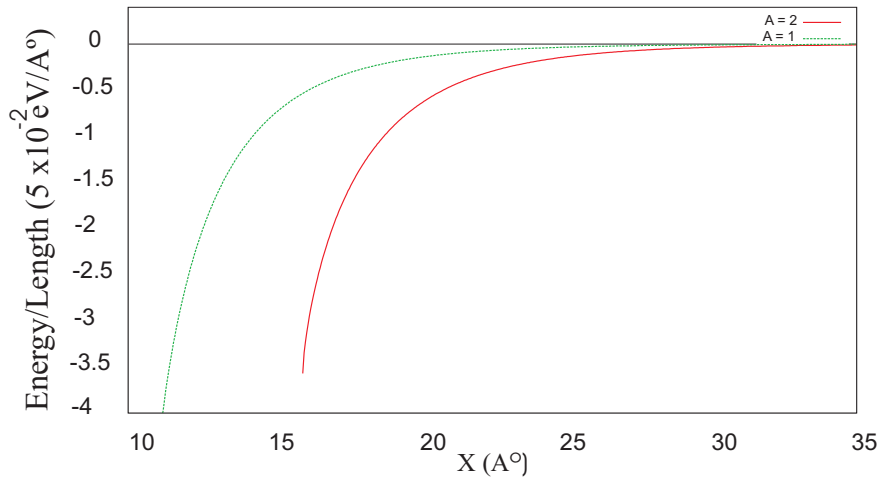}
  \caption{Interaction energy for the four-strand-rhombic case with $a=1$ (upper curve) and $a=2$ (lower curve). Distances $x$
  are measured in units of $1/ \mu \sim$ ${\cal O} (10)$ \AA, while lineal energy density units are estimated to be
  $5 \times 10^{-2}$~eV/\AA.}
\label{Both_Configs_Rhombus_Energy}
\end{figure}

\begin{figure}
 \centering
  \includegraphics[height=.3\textheight]{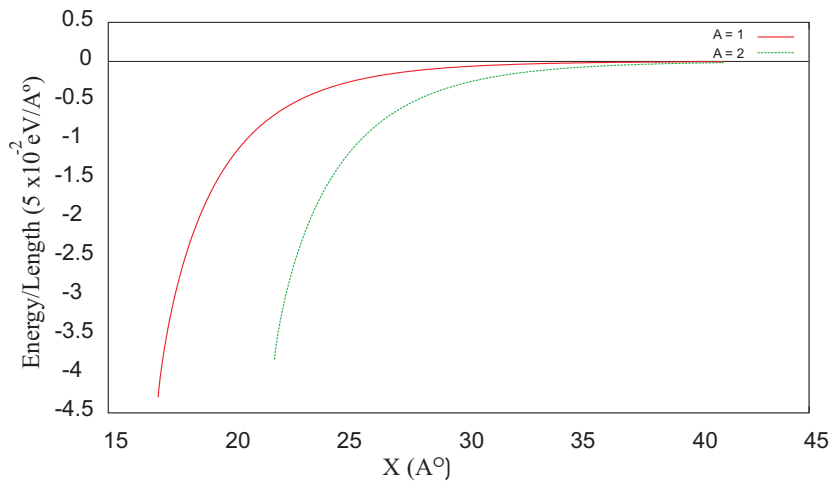}
  \caption{Interaction energy for the hexagonal lattice 19-strand configuration of Fig.~\ref{steps} with $a=1$ (upper curve) and $a=2$ (lower curve). Distances $x$ are measured in units of $1/ \mu \sim$ ${\cal O} (10)$ \AA, while lineal energy density units are estimated to be
  $5 \times 10^{-2}$~eV/\AA.}
\label{Two_19_String_Cases_Energy}
\end{figure}

Having established the above we have computed the energy for several interacting strands having various configurations. We present in Fig.~\ref{Both_Configs_Rhombus_Energy} and in Fig.\ref{Two_19_String_Cases_Energy} the results for four strands sitting at the vertices of a rhombus and for 19 strands arranged as in Fig.~\ref{steps}, respectively. Comparing these plots with that of the two strands interaction we clearly see that the attraction becomes stronger and acts on a larger range when the number of strands increases.

In the real case of aggregates it is always several DNA strands that interact, the two strands being only an idealization. Thus the fact that for 19 strands we find that (for $a=2$) the range of the force is in agreement with the typical values reported for DNA aggregates \cite{Hud} we take it as an indication of the validity of our hypothesis that the quantum Casimir energy could hold together the aggregates. Furthermore, this force is many-body in nature and the many-body effects are big, another reason for taking the two-body interaction only as an indication of the real phenomenon.

\begin{figure}
 \centering
  \includegraphics[height=.3\textheight]{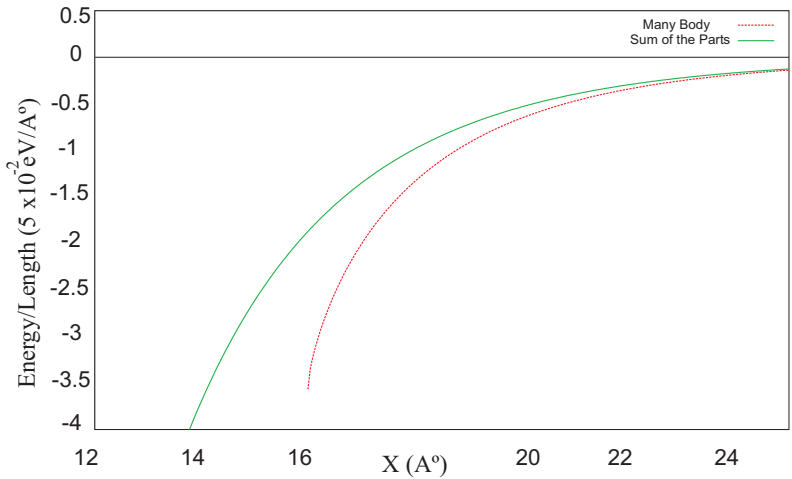}
  \caption{The lower curve is the interaction energy for the four-strand-rhombic configuration (many-body). The upper curve is what is obtained
  by summing-up the 6 two-body interactions. In both cases $a =2$. Distances $x$ are measured in units of $1/ \mu \sim$ ${\cal O} (10)$ \AA, while lineal energy density units are estimated to be $5 \times 10^{-2}$~eV/\AA.}
\label{4_Rhombus_Many_Body_vs_Sum_of_Parts}
\end{figure}

\begin{figure}
 \centering
  \includegraphics[height=.3\textheight]{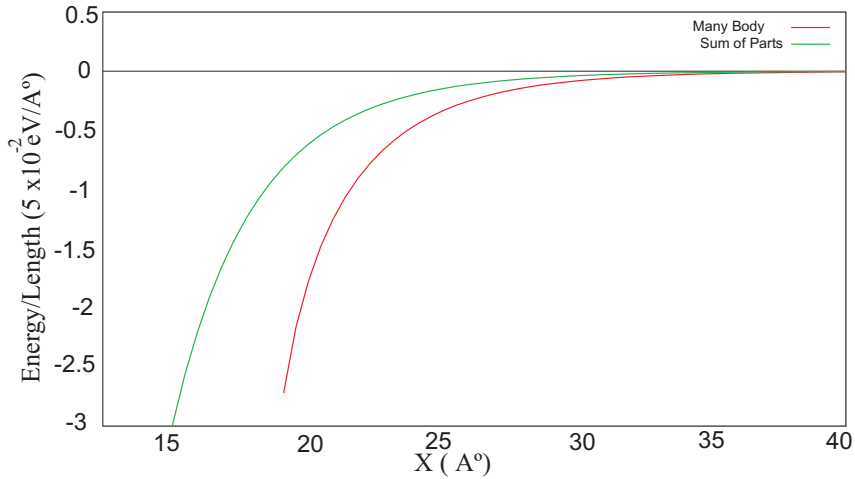}
  \caption{The lower curve is the interaction energy for the 7-strand configuration, i.e. for 7 strands sitting at the vertices and at the certer
  of a regular hexagon (many-body). The upper curve is what is obtained by summing-up the 21 two-body interactions. In both cases $a =2$. Distances $x$  are measured in units of $1/ \mu \sim$ ${\cal O} (10)$ \AA, while lineal energy density units are estimated to be $5 \times 10^{-2}$~eV/\AA.}
\label{2K0_7_String_Many_Body_vs_Sum_of_Parts}
\end{figure}

That the many body effects are strong we proved in our numerical calculations where we compared the $N$-body energy of Eq.~(\ref{ElndetNum}) with that
obtained by summing up $(N/2)(N-1)$  two-body interactions. The results for four and seven strands are shown in Fig.~\ref{4_Rhombus_Many_Body_vs_Sum_of_Parts} and Fig.~\ref{2K0_7_String_Many_Body_vs_Sum_of_Parts}, respectively, and they indicate that the effect
grows with $N$.

\section{Conclusions}\label{sec5}

The main result of this paper is to hint at quantum relativistic effects as viable candidates for the collapse of DNA strands
into aggregates (after the OM condensation has taken place) and for holding them together into stable condensates. Rather than
approaching the problem by setting-up from scratch a model based on quantum electrodynamics we built-up an effective model
based on the classical PB equation of electrostatics and included the DNA strands (modeled here as infinite lines or delta
functions) in the Boltzmann distribution, an instance that lead to a modified PB equation. The time-dependent fluctuations
we then studied are quantum in nature, propagate at the speed of light in the medium and give rise to a Casimir force that we studied in various settings. In particular, we focused our attention on the difficult problem of computing such interaction for several DNA strands and were able to overcome the analytical challenges with numerical calculations performed for a variety of cases, most of which with hexagonal symmetry
of arrangement (the typical situation reported in experiments).

The numerical calculations show many interesting features: the interaction is attractive and short range for the two-body case and is
many-body, the departures from the ``sum of two-bodies'' being important and growing with the number of strands; the distance at which our approximations stop working is also obtained as the value at which the determinant function vanishes; finally, the magnitude
and range of the interaction is such that it could explain the formation and stability of DNA aggregates, as a preliminary
comparison with reported data shows.

Our model is a primitive one and to make full contact with experiments we propose it as one part of the puzzle as it needs to be seen as one of the concurrence of \textit{three} mechanisms: (i) the zero-point quantum interaction, that gives the universal attraction (``glue''); (ii) the ``frustrated force'', that takes into account the detailed structure (finite size and helical architecture) and the active role of counterions (these alone seem to reproduce the helical architecture \cite{bruinsmapre}; (iii) water-related forces, that need to be included. That the interaction (\ref{Elndet}) \textit{alone} does not give the full picture is seen from the singularity at a value $\bar{x}$ such that the determinant becomes zero. If only this force were present the DNA-cation complexes would collapse to zero separation, but at very short distances we should include nonlinear corrections to this force itself, and, even leaving aside water-related forces, at short distances the Wigner crystal effects should become important. Our goal here was to \textit{single-out} the role of the zero-point quantum vacuum energy and we have shown that it is quite plausible that it plays a key role in the onset of the formation and in the follow-up stability of DNA aggregates.

It is pleasant to see that quantum effects might be essential for understanding an important biological problem (other quantum effects are important for enzyme catalysis \cite{Garcia-Viloca} or speculated to be important for neural activity  \cite{penrose}) as this might serve as a solid basis for a more general understanding of the role of quantum mechanics for life \cite{schroedinger}.

\section*{Acknowledgments}

A.I. benefitted from the many excellent talks and from the enlightening discussions ignited by Adrian Parsegian at the conference ``From DNA-inspired physics to physics-inspired biology'' held at ICTP, Trieste, in June 2009. Siddhartha Sen acknowledges the kind hospitality of the Institute for Particle and Nuclear Physics, Faculty of Mathematics and Physics of Charles University of Prague, where most of this work was carried out.

\end{document}